\documentclass[oldversion]{aa}
\usepackage{graphicx}
\usepackage{txfonts}
\begin{document}
   \title{Molecular gas in NUclei of GAlaxies (NUGA)}

   \subtitle{VIII The Seyfert 2 NGC~6574}

 \author{E. Lindt-Krieg
    \inst{1,2},
 A. Eckart\inst{1,3}, R. Neri\inst{2}, M. Krips\inst{4}, J.-U. Pott\inst{1,5}, S. Garc\'{\i}a-Burillo\inst{6}, F. Combes\inst{7}}

 \offprints{A. Eckart; eckart@ph1.uni-koeln.de}

 \institute{Universit{\"a}t zu K{\"o}ln, 1.Physikalisches Institut, Z{\"u}lpicher Strasse 77,  50937 K{\"o}ln, Germany
   \and
    Institut de Radio-Astronomie Millim{\'e}trique (IRAM), 300 rue de la Piscine, 38406 St. Martin-d'H{\`e}res, France 
\and
  Max-Planck-Institut f\"ur Radioastronomie, Auf dem H\"ugel 69, 53121 Bonn, Germany
\and
    Smithsonian Astrophysical Observatory (SAO),  Submillimeter Array (SMA)645, North A'Ohoku Place, 96720 Hilo, USA
\and
    W.M. Keck Observatory, 65-1120 Mamalahoa Hwy, Kamuela, HI 96743, USA
\and
 Observatorio Astron\'omico Nacional (OAN), Alfonso XII, 3, 28014 Madrid, Spain
\and
    Observatoire de Paris, LERMA, 61 Av. de l'Observatoire, 75014 Paris, France
             }
   \date{Received 17. Aug. 2007; accepted 2. Nov. 2007}

 \abstract
{Within the frame of the NUclei of GAlaxies (NUGA) project, we have determined 
the distribution and kinematics of the molecular gas within the central kpc
with high spatial resolution (100-150pc), for a sample 
of active galaxies. The goal is to study the gas-fueling mechanisms in AGN.

We present interferometric observations of $^{12}$CO(1-0) and $^{12}$CO(2-1) line emission 
from the Seyfert~2 galaxy NGC~6574, obtained with the IRAM Plateau de Bure Interferometer (PdBI). 
These data have been combined with 30m mapping data in these lines to correct for the flux 
resolved by the interferometer. At an angular resolution of 0.7$''$ ($\equiv$110pc), the $^{12}$CO(2-1)
emission is resolved into an inner disk with a radius of 300~pc.

The molecular gas in NGC~6574 is primarily distributed in four components:  nucleus, bar, 
spiral arms - winding up into a pseudo-ring - and an extended 
underlying disk component. For the overall galaxy host, we find a $^{12}$CO(2-1) to 
$^{12}$CO(1-0) line ratio of $\sim$0.3 indicative of cold or sub-thermally excited gas. 
For the nucleus, this ratio is close to unity, indicating emission from dense and warm 
molecular gas. Modeling the gas kinematics with elliptical orbits shows that the 
molecular gas in the differentially rotating disk of NGC~6574 is strongly influenced 
by the presence of  a stellar bar. The nuclear component shows an extension toward 
the southeast that may be an indication of the lopsidedness of the nuclear gas distribution.

We computed the gravity torques exerted from the stellar
bar on the gas, deriving the gravitational potential from near-infrared images,
and weighting the torques by the CO distribution. We find negative torques
for the gas inside the ring, since the gas aligned with the bar has
a slight advance phase shift, leading the bar. This means that gas is
flowing in towards the center, at least down to 400pc in radius,
which can explain the observed high nuclear gas concentration.
This concentration corresponds to a possible inner Lindblad resonance
of the bar, according to the measured rotation curve. The gas has been
piling up in this location quite recently, since no startburst is been observed yet.
}
	\authorrunning{E. Lindt-Krieg et al.}
	\titlerunning{Molecular gas in NUclei of GAlaxies (NUGA) VIII: NGC~6574}
   \maketitle
%

\section{Introduction}

The study of interstellar gas in the nuclei of galaxies is a fundamental for 
understanding nuclear activity through fueling the active galactic nuclei (AGN) 
and their relation to circum-nuclear star formation. 
Some studies claim that pure gaseous density waves (spirals, bars, warps, or lopsided 
instabilities) may be driving gas inflow towards the AGN 
(Heller $\&$ Shlosman 1994; Elmegreen et al. 1998; Regan $\&$ Mulchaey 1999). 
While such dynamical perturbations are responsible for the infall of gas on large scales, 
the processes responsible for removing angular momentum on small scales (sub-kpc) 
have not been understood very well yet. 

Different scenarios have been introduced trying to explain this phenomenon, such 
as nested bars (e.g., Friedli \& Martinet 1993, spirals e.g., Martini \& Pogge 1999,) 
warped nuclear disks (Schinnerer et al. 2000a,b), and lopsidedness or m=1 instabilities 
(Kormendy \& Bender 1999; Garc\'{\i}a-Burillo et al. 2000). 
Most of the gas within the central kiloparsec of spiral galaxies is in the molecular phase, 
while the atomic gas is deficient there. 

This makes CO lines the best tracers of nuclear gas dynamics in the nuclear interstellar medium. 
High-resolution (1$''$-2$''$) maps of the molecular component in the centers of galaxies 
are required to model these nuclei within the NUclei of GAlaxies (NUGA) project. 
The NUGA project aims at establishing a high angular resolution ($\approx$ 0.5$''$ - 1$''$) 
and high-sensitivity CO survey of twelve objects.  It covers the whole sequence of activity types: 
Seyfert 1, Seyfert 2, LINERs, and transition objects, and it comprises 12~objects as a core sample and a total of 
about 30 objects as an extended sample. The survey is being carried out with the IRAM Plateau de Bure 
mm-Interferometer (PdBI) in France, because 
it offers the best combination of sensitivity and 
resolution, both crucial for this project. 
A more detailed description of the NUGA project is given in Paper I 
by Garc\'{\i}a-Burillo et al. (2003). 

Previous surveys of molecular gas in active galaxies have been carried out by 
other groups (Heckman et al. 1989; Meixner et al. 1990; Vila-Vilaro et al. 1998; Sakamoto et al. 1999a,b), 
but have had insufficient spatial resolution (4-7$''$) to resolve the nuclear 
disk structures, or else were limited to small samples (Tacconi et al. 1997; Baker 2003). 
Besides case-by-case analyses and simulations of each object of the 
sample (Garc\'{\i}a-Burillo et al. 2003; Combes et al. 2004; 
Krips et al. 2005; Garc\'{\i}a-Burillo et al. 2005), 
the collected data of the NUGA project also provides an initial statistical basis 
for studying different mechanisms that may be responsible for gas flow toward the nuclei 
or may account for further accretion inward.

This paper describes the distribution and dynamics of molecular gas in 
NGC~6574, one of the galaxies belonging to the NUGA survey. 
NGC~6574 is a Seyfert 2 galaxy of Hubble type SB(s)bc, at a distance of 33 Mpc.
This results in a spatial scale of $\sim$160~pc per arcsecond. 
Its main characteristic is a symmetric structure with two spiral arms 
to the north and south of the center, nested on a small bar within the central 8$''$.
Kotilainen et al. (2000) present high spatial-resolution, near-infrared broad-band JHK and 
Br$\gamma$ 2.166 $\mu$m and H$_2$ 1-0 S(1) 2.121 $\mu$m emission line images of the circum nuclear star-formation ring.
The overall near-infrared and radio morphologies (see Laine et al. 2006) generally agree with each other. 
The observed H$_2$/Br$\gamma$ ratio indicates that the main excitation mechanism of the molecular gas is 
UV radiation from hot young stars. Shocks are likely to contribute only in a few regions. To explain 
the NIR data, Kotilainen et al. (2000) prefer the model of an instantaneous burst of star formation with an upper 
mass cutoff of M$_u$ = 100 M$_{\odot }$ occurring $\sim$6-7~Myr ago. 
An analysis of the molecular gas distribution at an angular resolution of about 4$''$ in the $^{12}$CO(1-0) 
line is presented by Sakamoto et al. (1999ab). They find a prominent CO peak of 1~kpc diameter at the center 
of the galaxy on a smooth gas disk with two gas/dust lanes that extend about 2~kpc from the central peak 
to east and west at a PA of $\sim$105$^\circ$. From the tips of this gaseous bar, spiral arms start 
to form a ring-like structure.

In this paper we analyze data obtained with the IRAM PdBI and the 30~m single-dish telescope in the 
$^{12}$CO(1-0) and $^{12}$CO(2-1) lines. The angular resolution we obtained in the combined data set is 
about 2$''$ and 0.7$''$. We describe the molecular gas distribution and kinematics with 
elliptical bar orbits, to account for the influence of the stellar bar potential
in the nucleus of NGC~6574. 
In Section 2 we describe the CO observations and in 
Section 3 we present the first results. 
The kinematic model and its results are discussed in Section 4.  
Gravity torques and the nature of gas flows are derived in Section 5, 
while Section 6 concludes.


\section{Observations and data reduction}

In this section we present the $^{12}$CO observations with the IRAM 30m single dish, 
with the IRAM PdBI, including the short-spacing correction.

\subsection{$^{12}$CO observations with the IRAM 30m single dish}

We performed IRAM 30m observations in a $7 \times 7$ raster pattern with 10$''$ spacing
in August 2005, in order to add the short spacings
and to estimate the flux density filtered out by the interferometric observations.
We used 4 SIS receivers to observe simultaneously at the redshifted
frequencies of the $^{12}$CO(1-0) and the $^{12}$CO(2-1) lines, thereby assuming the LSR
velocity of NGC6574 of 2282 km~s$^{-1}$. For each line, a 4~MHz channel filter bank was used with a velocity 
resolution smoothed to about 10 km~s$^{-1}$ in order to improve the signal-to-noise ratios of the individual spectra. 
The 4~MHz filter bank consists of nine units. Each unit has 256 channels which covers a total bandwidth of 1~GHz. 
At 115~GHz and 230~GHz, the HPBW is 21.5$''$ and 10.75$''$, respectively. The parameters for the two observing 
runs are summarized in Table~\ref{tab:1}. We used the wobbler switch mode with a 6~s cycle and a throw of 240$''$. 
With the forward efficiency $F_{eff}$, the main-beam efficiency is 
$B_{eff}$=$F_{eff}$$\times$$T_{\rm A}^*/T_{\rm mb}=0.74$ 
at 115~GHz and 
0.53 at 230~GHz. The typical system temperature varied between 200 and 450~K at both frequencies. The pointing 
was checked regularly on continuum sources and the accuracy was consistent to 3$''$ rms achieved by Greve et al. (1996). 
The total bandwidth available was 512~MHz at 115~GHz and 230~GHz, corresponding to 1330 km~s$^{-1}$ and 666 km~s$^{-1}$, 
respectively. The data reduction and analysis were done with the GILDAS package (e.g. Guilloteau \& Lucas 2000). 
For the short-spacing correction and further investigations, only the 4~MHz filter bank was used.

\begin{table}[!h]
\centering
\begin{tabular}{lcc}
\hline
\hline
Parameters & CO(1-0) & CO(2-1) \\
\hline
central coordinates & \multicolumn{2}{c}{RA = 18$^{h}$11$^{m}$51.3$^{s}$, DEC = +14$^{\circ}$58$^{'}$52.1$^{''}$} \\
backends & \multicolumn{2}{c}{4~MHz filter bank/autocorrelator}\\
redshift $z$ & \multicolumn{2}{c}{0.007662}\\
sky frequency [MHz] & 114.400 & 228.798\\
beam size [arcsec] & 21.50 & 10.75\\
beam efficiency $B_{eff}$ & 0.739 & 0.527\\
forward efficiency $F_{eff}$ & 0.95 & 0.91\\
total obs. time [sec] & 14994 & 14994\\
Nr. of positions & 49 & 49\\
Nr. of scans &  294 & 294\\
\hline
\hline
\end{tabular}
\caption[Parameters of observing run for IRAM 30m telescope]{Parameters of the observing run with IRAM 
30m single dish. 
}
\label{tab:1}
\end{table}

\subsection{$^{12}$CO observations with IRAM PdBI}

We observed the emission of the J=1-0 and J=2-1 lines of $^{12}$CO in a single 
field centered at the radio position of the AGN 
(i.e., $\alpha_{2000}$=18$^{\rm h}$11$^{\rm m}$ 51.3$^{\rm s}$and $\delta_{2000}$=14$^{\circ}$ 58'52.1$''$)
in NGC~6574 with the IRAM Plateau de Bure interferometer (PdBI) in April 2002 (D configuration), 
in November 2003 (C configuration) and February 2004 (A and C configurations). The six 15~m antennae were 
equipped with dual-band SIS receivers. The spectral correlators were centered at 114.400~GHz and 228.798~GHz,
respectively, with three correlator units covering a total bandwidth of 400~MHz at each frequency. The difference between 
LSR and heliocentric velocities is 12 km~s$^{-1}$, therefore the observations were centered on 
$v_{\rm hel} = 2282$ km~s$^{-1}$. The correlator was regularly calibrated by a noise source inserted in the IF system.

The bandpass calibrator was 3C 273 or MWC349, while the phase and amplitude calibration were performed on 
the nearby quasars 1923+210 and 1749+096, depending on the data set. Flux densities were calibrated 
relative to CRL618 and MWC349. The data were phase calibrated in the antenna-based mode. The frequencies were 
centered on the redshifted $^{12}$CO(1-0) line in the USB at 3 mm and on the redshifted $^{12}$CO(2-1) line 
in the LSB at 1 mm. For each line, the total bandwidth was 400~MHz and the spectral resolution 1.25~MHz. 
The integration time for the central pointing amounts to $\sim $16 hrs on the source. The water vapor ranged
 between 4 and 10 mm (i.e., opacities of $\sim $0.2-0.3) resulting in system temperatures of approximately 200-300 K on average.

The flux densities of the primary calibrators were determined from IRAM measurements 
and taken as input for deriving the absolute flux density scales for our 
visibilities, estimated to be accurate to 10\%. 
The parameters for the four observing runs are summarized in Table~\ref{tab:2}.

The data reduction and mapping was performed using the GILDAS software. 
Data cubes with $512 \times 512$ spatial pixels (0.25$''$/pixel) were created with velocity planes 
separated by 5~km~s$^{-1}$ (for PdBI alone, and by~10 km~s$^{-1}$ for the short-spacing corrected data). 
The cubes were cleaned with the Clark (1980) method for $^{12}$CO(1-0) data and with 
MX for $^{12}$CO(2-1) data. 
As synthesized clean beams we used a 2.48 $''$ $\times$ 1.69$''$ Gaussian 
with PA = 26$^{\circ }$ at 3~mm and a 0.93 $''$ $\times$ 0.5$''$ Gaussian 
with PA = 12$^{\circ }$ at 1~mm. The rms noise levels in the cleaned maps (at 5~km~s$^{-1}$ 
velocity resolution) are 3~mJy/beam and 6~mJy/beam for the $^{12}$CO(1-0) and $^{12}$CO(2-1) 
lines, respectively. No continuum emission was detected toward NGC~6574. 
The maps were corrected for primary beam attenuation. 
\begin{table}[!h]
\centering
\begin{tabular}{lcc}
\hline
\hline
Parameters & CO(1-0) & CO(2-1) \\
\hline
central coordinates & \multicolumn{2}{c}{RA = 18$^{h}$11$^{m}$51.3$^{s}$, DEC = +14$^{\circ}$58$^{'}$52.1$^{''}$} \\
redshift $z$ & \multicolumn{2}{c}{0.007662}\\
sky frequency [MHz] & 114.400 & 228.798\\
beam size [arcsec] & 2.48$\times$1.69 & 0.91$\times$0.48\\
position angle [degree] & 26 & 12\\
\hline
\hline
\end{tabular}
\caption[Parameters of observing run for PdBI]{Parameters of the observing run with IRAM PdBI.}
\label{tab:2}
\end{table}

\subsection{Short-spacing correction}

Short spacings were included using the SHORT-SPACE task in the GILDAS software. To find the best
compromise between good angular resolution and complete restoration of the missing extended flux, the 
weights attached to the 30m and PdBI %
\begin{figure}[!h]
   \centering
   \rotatebox{0}{\resizebox{7.5cm}{!}{\includegraphics{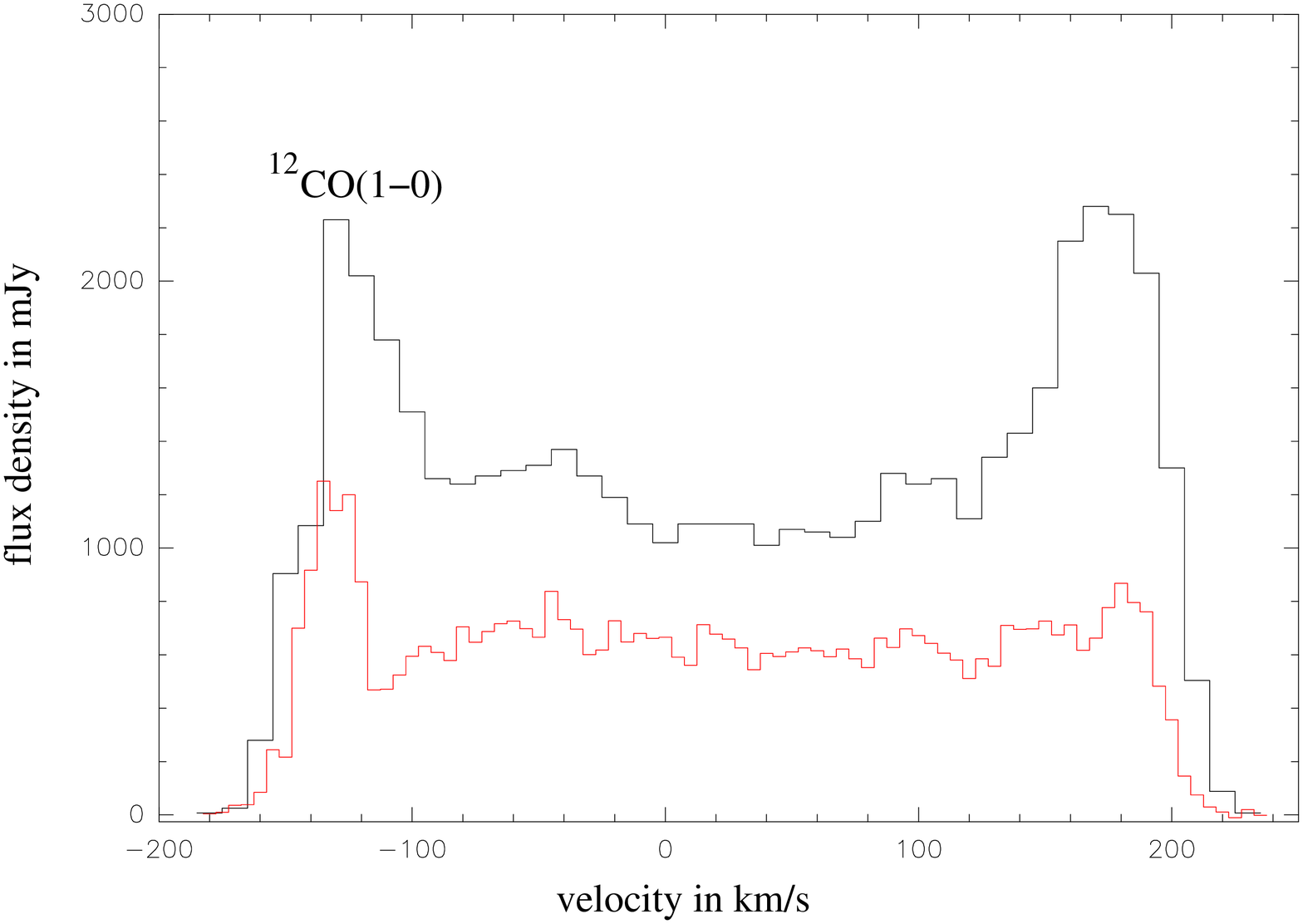}}}
   \rotatebox{0}{\resizebox{7.5cm}{!}{\includegraphics{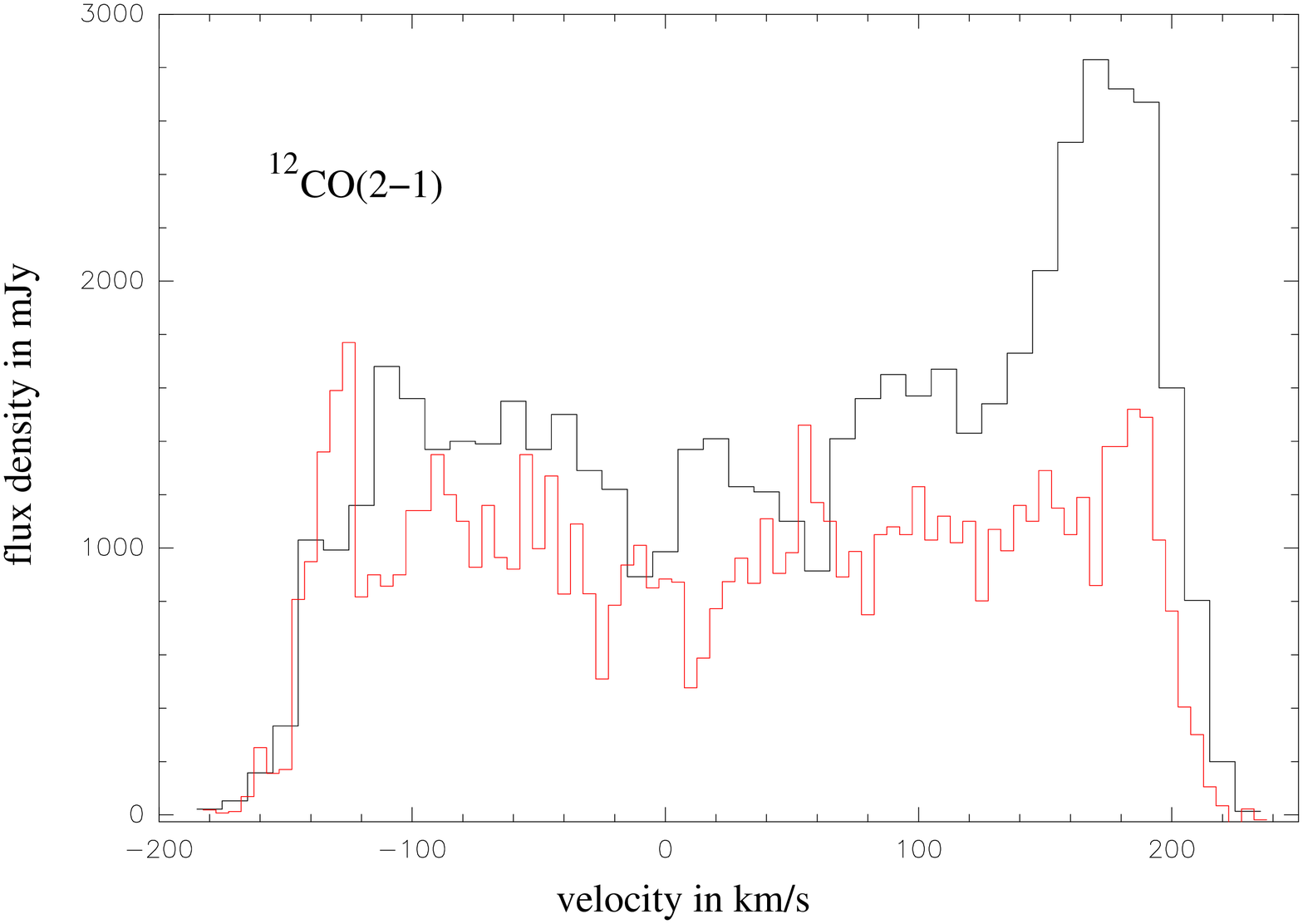}}}
   \caption{
The $^{12}$CO(1-0) and $^{12}$CO(2-1) line intensity derived from
the PdBI data set (red) and the combined PdBI+30m data set (black).
The spectra combine the information shown in the channel map in Fig.~\ref{Fig:1}.
The comparison shows the larger amount of extended emission for the $^{12}$CO(1-0) line
that is resolved by the interferometer. Both sets of spectra also demonstrate the 
asymmetry in flux density between the red and blue shifted part of the spectrum.
The red side of the spectrum is brighter mostly in its extended emission.
}
   \label{Fig:0}
\end{figure}
\begin{figure*}
\centering
\includegraphics[angle=-90, width=\textwidth]{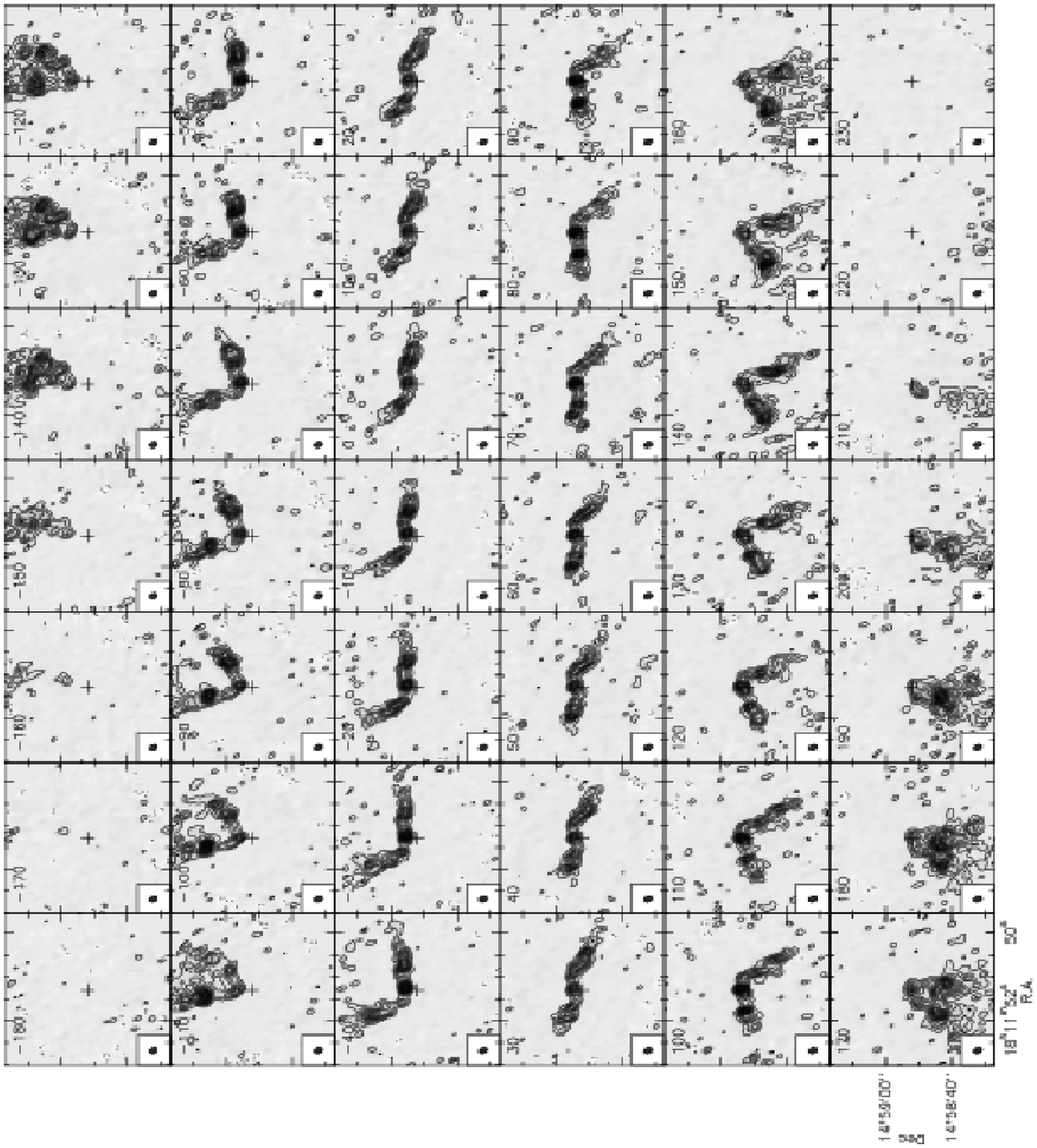}
\caption{$^{12}$CO(1-0) velocity-channel maps observed with the PdBI and corrected through short-spacing 
by the 30m single dish in the nucleus of NGC~6574. 
The clean beam resulting from the observations is 2.48$''$ $\times$1.69$''$ at PA=26$^{\circ}$. 
It is plotted as a filled ellipse in the bottom left corner of each panel.
We show a field of view of 55$''$. The phase-tracking center is indicated by a cross at $\alpha_{2000}$ = 18$^{h}$11$^{m}$51.3$^{s}$ 
and $\delta_{2000}$ = +14$^{\circ}$58$^{'}$52.1$''$. Velocity channels are displayed from $v=$ -180 km~s$^{-1}$ 
to $v=$ 230 km~s$^{-1}$ in steps of 10 km~s$^{-1}$. Velocities are in LSR scale. Contour levels are 0.42$\sigma$, 
0.84$\sigma$ to 4.8$\sigma$ in steps of 0.42$\sigma$, where the 1-$\sigma$ rms is $\sigma$=2.39 mJy~beam$^{-1}$.}
        \label{Fig:1}
   \end{figure*}
data were varied. 
At the end, a factor of two was applied to the weights of the 30m data enabling us to 
recover more than 40\% of the missing flux and to retain the angular resolution of 2.48$''$ $\times$ 
1.69$''$ at PA of 26$^{\circ}$ for $^{12}$CO(1-0) emission and 0.93$''$ $\times$ 0.5$''$ at PA of 
12$^{\circ}$ for $^{12}$CO(2-1) emission. 
The combined data sets were written to visibility tables, 
converted to maps using standard reduction procedures, and then deconvolved using the Clark algorithm. 
The weights were adjusted to get the same mean weights in the single-dish data as in the interferometer 
data in the u-v range of $1.25~D/\lambda$ to $2.5~D/\lambda$(D=15 m). 
All figures presented in this paper are done with short-spacing corrected data.

\begin{table}[!h]
\centering
\begin{tabular}{lcc}
\hline
\hline
Line  & I$_{CO}$ & L$_{CO}$   \\
      & Jy~km$^{-1}$  & K~km$^{-1}$~pc$^{2}$           \\
\hline
$^{12}$CO(1-0) &    &   \\
PdBI           & 223$\pm$~9   & 6.43$\times$10$^8$  \\
PdBI + 30m     & 485$\pm$15  & 13.5$\times$10$^8$  \\
$^{12}$CO(2-1) &    &   \\
PdBI           & 360$\pm$11   & 2.49$\times$10$^8$  \\
PdBI + 30m     & 540$\pm$17   & 3.69$\times$10$^8$  \\
\hline
\hline
\end{tabular}
\caption{Line intensities and CO luminosities integrated
over the entire linewidth of the full $\Delta$v=370~km~s$^{-1}$ 
and the extent of the combined PdBI+30m map after correcting corrected for primary beam
response (typical errors are of the order of 20\%).
}
\label{tab:4}
\end{table}

The line intensities and CO luminosities integrated
over the entire linewidth of the full $\Delta$v=370~km~s$^{-1}$ 
extent of the combined PdBI+30m map after correcting for primary beam
response are listed in Table~\ref{tab:4}. 
Since these maps contain the entire galaxy, the corresponding 
quantities give information on the overall molecular excitation in NGC~6574. 
The difference between the integrated quantities for the $^{12}$CO(1-0) 
line emission already indicate the presence of an extended disk 
component that is resolved out by the interferometer (see Fig.~\ref{Fig:0}).
\begin{figure*}
\centering
\rotatebox{0}{\resizebox{17.0cm}{!}{\includegraphics{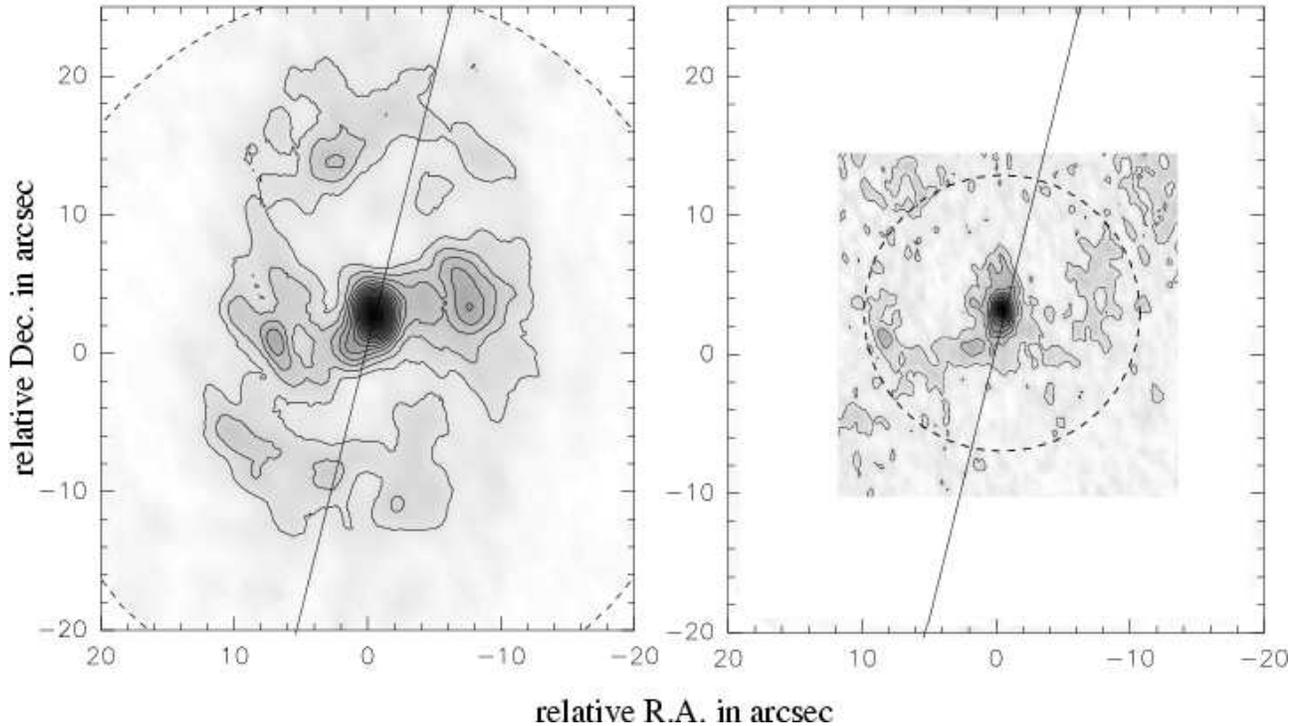}}}
\caption{Primary beam corrected integrated intensity maps for the 
 $^{12}$CO(1-0) ({\it left}) and  $^{12}$CO(2-1) ({\it right}) line emission.
The straight lines at PA = 165$^{\circ}$ indicate the position of the major axis of 
NGC~6574 along and perpendicular to which we obtain position-velocity diagrams.
The $^{12}$CO(2-1) map is shown in the same frame as the $^{12}$CO(1-0) map.
The dashed circles outline the central regions that approximately
include the FWHM areas of the PdBI primary beams at the two frequencies. 
For larger radii the map is too noisy. 
Inside that radius the nuclear emission as well as emission from
the bar can be detected.
This includes the tips of the bar at which the spiral arms start.
Contour lines are 
1, 2, 3, 4, .... 16 Jy~km/s~/beam 
for $^{12}$CO(1-0)
and 
1.8, 3.6, 5.4, 7.2, ... 15 Jy~km/s~/beam 
for $^{12}$CO(2-1) 
line emission map, with beam sizes given in Table~2.
}
\label{Fig:2}
\end{figure*}

\section{Results}

In this section we first describe the morphology of NGC~6574 and show 
its kinematics, then estimate the gas masses and related parameters.

\subsection{Morphology of the CO emission}

In Fig.~\ref{Fig:1} we show the velocity-channel maps of the $^{12}$CO(1-0) line emission 
in the central region of NGC~6574, with a velocity range of 
410 km~s$^{-1}$ and a velocity resolution of 10 km~s$^{-1}$. 
The typical 'butterfly' structure visible in the iso-velocity maps in Fig.~\ref{Fig:5} 
is the signature of a spatially resolved rotating disk. 
In the central 4$''$-5$''$ the location of the nuclear peak emission in 
the $^{12}$CO emission also moves from the northern part of the channel 
map at the velocity of -170 km~s$^{-1}$ to the southeast at 
the velocity of 220 km~s$^{-1}$ with respect to the phase-tracking center of the map.
The displacement of the nuclear component as a function 
of velocity shows that it is resolved, and it implies the rotation of the nuclear gas 
around the very center of NGC~6574.
In the $^{12}$CO(1-0) integrated intensity map shown in Fig.~\ref{Fig:2} (left), 
the molecular gas distribution reveals a regular ring-like structure, 
with a mean 
angular diameter of $\sim$50$''$ 
and the peak of the emission in the central 
4$''$-5$''$. The integrated map was produced by averaging the channel maps 
over a total velocity width of 415 km~s$^{-1}$ (from -180 to 235 km~s$^{-1}$). 
For the $^{12}$CO(1-0) emission, the ring consists of some molecular cloud 
complexes that appear to be connected to each other. 
They form the two spiral arms of NGC~6574 that start at a PA of $\sim$100$^{\circ}$
and a projected radius of 
$r\sim$1.1~kpc (1.55~kpc deprojected with an inclination of 45$^{\circ}$ - see below), 
near the end points of a prominent nuclear bar and extend out to radii of about $r\sim$2.4 kpc 
towards the north and south, approximately along the major axis. 
A comparison to NIR images (Kotilainen et al. 2000) reveals that the position of 
the central, resolved CO source coincides to within less than 0.25$''$
with that of the AGN.
From the $^{12}$CO(1-0) map in Fig.~\ref{Fig:2} (left) we estimate a gas bar PA of about 
115$^\circ$ $\pm$5$^\circ$, taking into account that the line emission on the western and
eastern sides of the bar appear to be displaced by about 1$''$-2$''$ to the south and north, respectively.
Our value of 115$^\circ$ is in reasonably agreement with the value of 105$^\circ$ 
(towards the western tip of the bar and derived from a lower resolution map) 
given by Sakamoto et al. (1999ab). It also apparently agrees with the NIR maps presented
by Kotilainen et al. (2000).
The NIR stellar bar has a PA of 105$^\circ$, which means that our 
high-resolution CO maps are able to detect a slight advance phase shift between
the stellar and gas bars, the latter leading by about 10$^\circ$.
Kotilainen et al. (2000) also interpret the elongation of the nuclear
component as a possible nuclear bar, with an orientation of PA= 150$^\circ$,
however, this feature is not certain and could be due to projection
effects, as the PA of the major axis is 165$^\circ$. 
The CO spiral arms, starting at the end of the bar,
 mostly lie at the trailing edges of the nuclear bar. 
This particular geometry may determine the feeding budget for the gas 
in this region. The spiral arms and the central source are connected by a 
molecular gas bridge - the small bar.
Contrary to $^{12}$CO(1-0), the $^{12}$CO(2-1) integrated emission is detected 
mainly inside the central component within the radius of 4$''$-6$''$, as shown 
in Fig.~\ref{Fig:2} (right). The maximum of the emission is concentrated at the center as well. 
While our $^{12}$CO(2-1) map lacks sensitivity outside the central 500~pc of the disk, 
it does give a sharp image of the molecular gas distribution in the vicinity of the AGN. 
Also the tips of the nuclear bar in the eastern and western parts of the galaxy
are indicated.

\subsection{Kinematics of the molecular gas disk}

\begin{figure*}
\centering
	\rotatebox{000}{\resizebox{18.0cm}{!}{\includegraphics{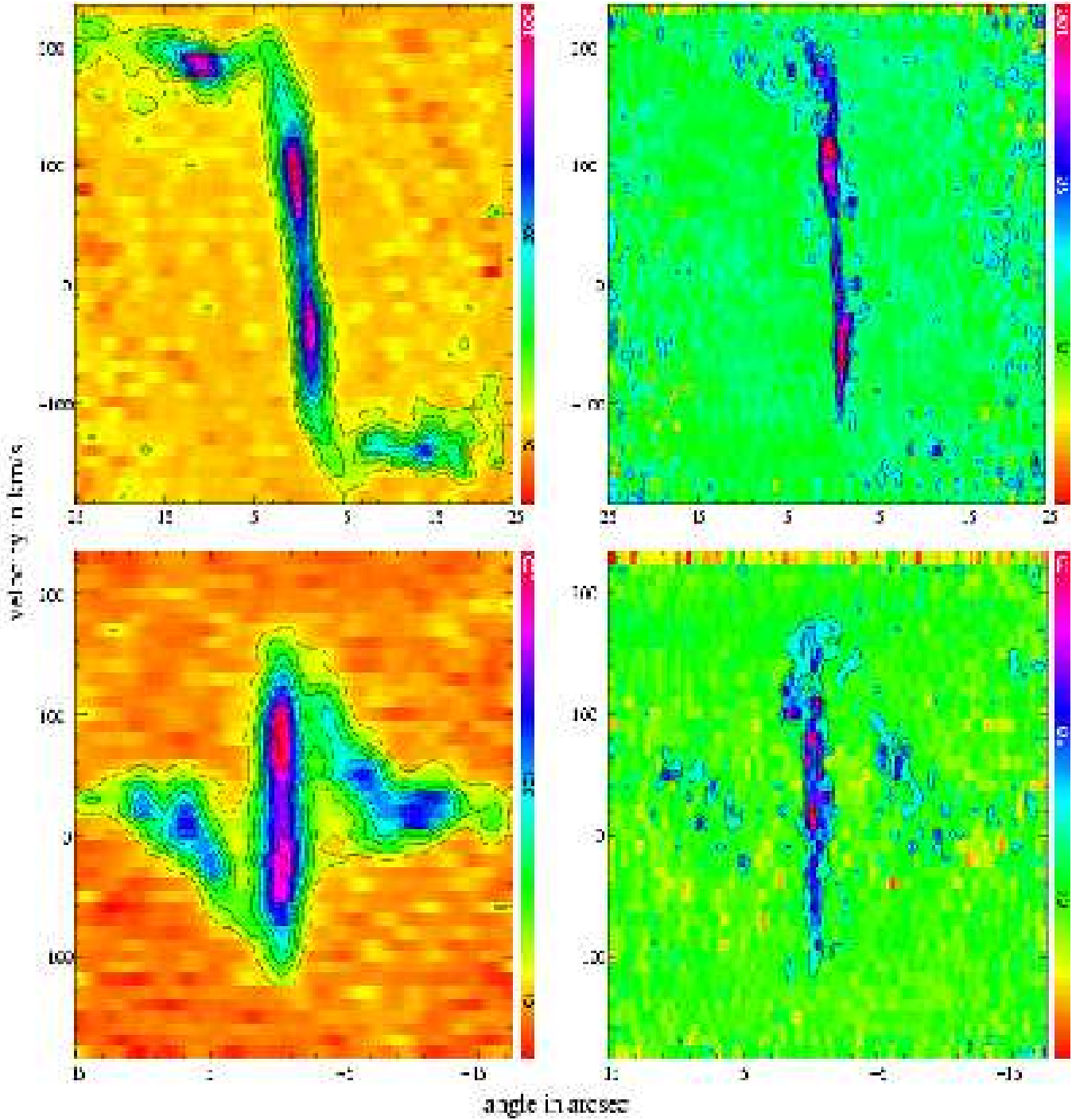}}}
  \caption{Position-velocity diagrams of $^{12}$CO(1-0) and $^{12}$CO(2-1) along the major axis 
at 165$^{\circ}$ ({\it {upper panels}}) and minor axis at 75$^{\circ}$ ({\it {lower panels}}). 
For the top and bottom diagrams, the left side corresponds to the approximate south and east 
directions, respectively. 
The color bar gives the relative peak intensities in percent, with 0, 50, and 100\% levels indicated
just below the corresponding marks or the top of the bar.
From the top left to the bottom right graph the contour lines are equidistantly 
spaced in 7, 17, 10, and 17\% of the peak intensity, respectively.
}
        \label{Fig:4}
   \end{figure*}

The kinematics of the molecular gas are dominated by its motion in a rotating disk.
In addition we see the influence of spiral arms and a bar potential. 
In Fig.~\ref{Fig:4} we show the position-velocity (p-v) plot along the major 
axis at PA = 165$^{\circ}$ and along the minor axis at PA = 75$^{\circ}$.
By comparing the 30m and PdBI spectra extracted from the corresponding maps at the same positions,
we find that there is some excess extended emission 
(as also suggested by Fig.~\ref{Fig:0}),
especially in the southern part of NGC~6574. 
Therefore both the PdBI and the 30m data indicate that the molecular gas emission is asymmetric 
with respect to $v_{\rm sys}$, so the CO emission is preferentially redshifted
(see Fig.~\ref{Fig:0}).
Our results agree well with the previous finding by Sakamoto et al. (1999ab), which were 
obtained at a lower resolution.

The position velocity diagram along the major axis shows the rotation curve that still has to be 
corrected for the inclination of the galactic disk. 
For both lines of $^{12}$CO emission, the results are similar; however, 
differences between them can be found in the central 2$''$. 
The p-v diagram of the $^{12}$CO(1-0) line resolves the disk and nuclear region of the galaxy.
There is some evidence of $^{12}$CO(2-1) line emission at the very central position, although
the diagram clearly shows that the emission at the systemic velocity is suppressed with respect to
the emission arising at 0.5$''$ to a 1$''$ separation from the nucleus. 
This expresses itself in 
two emission clumps neighboring the central position and systemic velocity in the 
$^{12}$CO(1-0) p-v diagram. This could hint at the existence
of a small nuclear ring. The minor axis feature is a clear indication 
of non circular motion, expected in a bar, supporting the nuclear
bar hypothesis.
In the $^{12}$CO(2-1) p-v diagram this contrast between the clumps is also visible on
the major axis and reduced on the
minor axis. However, the $^{12}$CO(2-1) map has less sensitivity, due to its increased
spatial resolution.

\begin{figure*}
\centering
\includegraphics[angle=-0, width=\textwidth]{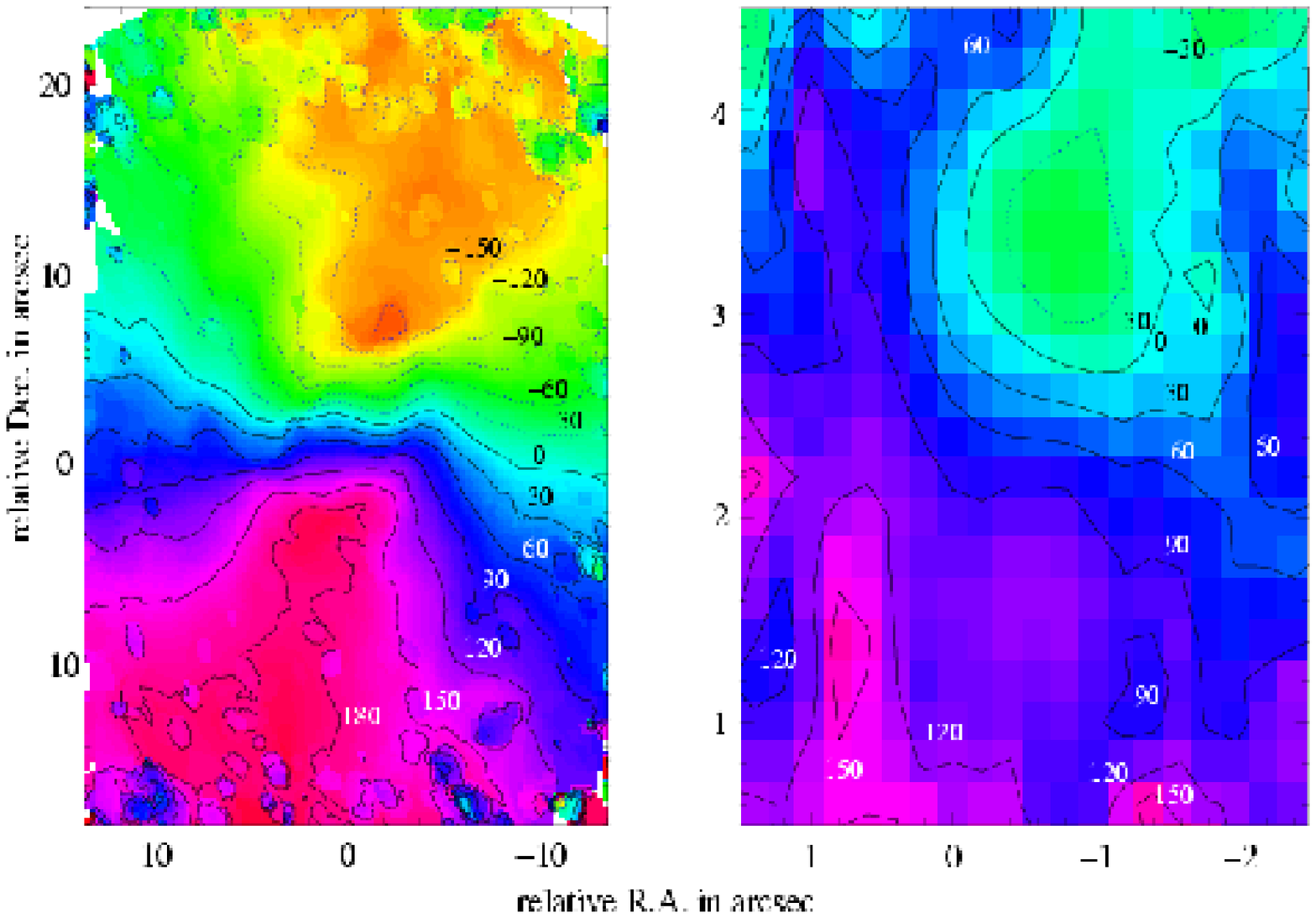}
      \caption{Iso-velocity diagram of $^{12}$CO(1-0) ({\it {left panel}}) and $^{12}$CO(2-1) 
emission ({\it {right panel}}). The $^{12}$CO(2-1) emission is observed only in 
the central part (compare with left panel). The innermost solid contour line 
is at the systemic velocity and the velocity steps are 30~km~s$^{-1}$.}
        \label{Fig:5}
   \end{figure*}

The distribution of the $^{12}$CO emission, as well as the iso-velocity maps, also indicates
a more complex molecular gas kinematics in the center of the galaxy.
The iso-velocity map of $^{12}$CO(1-0) and $^{12}$CO(2-1) line emission were made from maps at the full angular 
resolution of 2.48$''$$\times$1.62$''$ and 0.93$''$$\times$0.5$''$, respectively.
In both cases we used equally spaced velocity channels with a velocity range from 
-170 to 220 km~s$^{-1}$  (see Fig.~\ref{Fig:5}) with the central channel centered on the systemic velocity.
The maps show the increase in velocities from north to south in both lines. 
The iso-velocity lines of the central 10$''$ are parallel to each other but tilted
by about 20$^{\circ}$ with respect to 
those of the outer disk sections of NGC~6574. This is a clear indication of the presence of spiral arms
at the location of the bends between the iso-velocity contours and the influence of a barred potential
that expresses itself in an overall tilt of the velocity field with respect to that of the molecular disk.
While the $^{12}$CO(1-0) line shows the velocity field for the whole galaxy (disk + nucleus), the 
$^{12}$CO(2-1) line emission only allows us to derive the velocity field for the central 4$''$ $\times$4$''$ region.
Here the overall orientation of the iso-velocity lines is similar to the $^{12}$CO(1-0) but shown at a higher
angular resolution - hence the influence of the barred potential is more pronounced.

\subsection{Molecular gas masses}

The velocity-integrated $^{12}$CO(1-0) flux within the $\sim$50$''$ primary beam of the combined IRAM PdBI/30m data
is used to estimate molecular hydrogen masses. 
We use 
$I'_{CO}$[K~km~s$^{-1}$] = 0.12$\times$10$^{25}$$\cdot$$I_{CO}$[Jy~km~s$^{-1}$]$\cdot \nu^{-2}$[Hz]$\cdot \theta^{-2}$[$''$],
with the observing frequency $\nu$ and the beam size of the corresponding map $\theta$.
The column density of molecular hydrogen can be derived from $I'_{CO}$ by assuming the validity of an average 
conversion factor $X$ = $N(H_{2})/I'_{CO}$ = 2.2$\times$10$^{20}$ [cm$^{-2}$K~km~s$^{-1}$] 
(see e.g. Solomon \& Barrett 1991):

\begin{equation}
\centering
N(H_{2}) = X[\frac{\textnormal{cm$^{-2}$}}{\textnormal{K~km~s$^{-1}$}}] \cdot I'_{CO}[\textnormal{K~km~s$^{-1}$}]~~.
\end{equation}

To estimate the molecular gas mass, the column density $N(H_{2})$ then needs to be
multiplied with an effective projected area $A$ over which the CO line emission is distributed:

\begin{equation}
\centering
M_{H_{2}}[M_\odot] = 1.51 \times 10^{-14} \cdot N(H_{2})[\textnormal{cm$^{-2}$}]\cdot A[\textnormal{kpc$^{2}$}]~.
\end{equation}

\noindent
Including the relative contribution of helium and and neutral atomic hydrogen, the total gas mass 
contained in the disk of NGC~6574 can be estimated via
\begin{equation}
\centering
M_{gas}[M_\odot] =  
1.36 \cdot (M_{H_{2}}+M_{H_{I}})~,
\end{equation}
with helium mass contained in the factor of 1.36.

The total dynamical mass of NGC~6574 is derived from the virial theorem by using 
the inclination-corrected circular velocity for a given radius via 
M$_{dyn}$[M$_{\odot}$] = 232 $\times$ v$^2_{rot}$(r)[km s$^{-1}$] $\times$ r[pc]. 
The error of the Keplerian dynamical mass due to the flatness 
of the mass distribution is at most 30\% for an exponential disk 
(Binney \& Tremaine 1987).
All results are summarized in Table~\ref{tab:3}, separated into 
two components: the nucleus, which contains the flux of the inner 4$''$,
and the disk, containing the residual flux within the 50$''$ of the 
integrated map.
\begin{table}[!h]
\centering
\begin{tabular}{lcc}
\hline
\hline
Parameters & PdBI & PdBI+30m \\
\hline
\multicolumn{3}{l} {nucleus:}\\
$I_{CO_{10}} [Jykms^{-1}]$ & 17.57 & 20.30\\ 
$I'_{CO_{10}} [Kkms^{-1}]$ & 1.45 & 1.86 \\
$N(H_2)[cm^{-2}]$ & 3.18$\times$10$^{20}$ & 4.09$\times$10$^{20}$ \\
$M_{H_{2}}^{CO_{10}}[M_\odot]$ & 1.05$\times$10$^{8}$ & 1.18$\times$10$^{8}$ \\
$M_{H_{2}+He}^{CO_{10}}[M_\odot]$ & 1.43$\times$10$^{8}$ & 1.60$\times$10$^{8}$\\
\multicolumn{3}{l} {disk:}\\
$I_{CO_{10}} [Jykms^{-1}]$ & 223 & 485\\ 
$I'_{CO_{10}} [Kkms^{-1}]$ & 0.83 & 2.04  \\
$N(H_2)[cm^{-2}]$ & 1.86$\times$10$^{20}$& 4.47$\times$10$^{20}$ \\
$M_{H_{2}}^{CO_{10}}[M_\odot]$ & 6.07$\times$10$^{7}$ & 1.29$\times$10$^{8}$ \\
$M_{H_{2}+He}^{CO_{10}}[M_\odot]$ & 8.26$\times$10$^{7}$ & 1.74$\times$10$^{8}$\\
$M_{gas}^{CO_{10}}[M_\odot]$ & 2.04$\times$10$^{9}$ & 2.13$\times$10$^{9}$\\
\multicolumn{3}{l} {dynamical mass:}\\
$M_{dyn}^{CO_{10}}[M_\odot]$ & & 3.5 $\times$ 10$^{10}$  \\
\hline
$M_{H_{I}}[M_\odot]$ & 1.44$\times$10$^{9}$ & (Mirabel \& Sanders, 1988)\\
\hline
\hline
\end{tabular}
\caption[Results of IRAM PdBI and PdBI+30m telescope]{Results of IRAM PdBI and PdBI+30m telescope.}
\label{tab:3}
\end{table}
This central source contains a molecular gas mass of $\sim$ 10$^{8}$ M$_{\odot }$. 
Estimated from the kinematics revealed by the $^{12}$CO(1-0) line emission,
the total dynamical mass of NGC~6574 contained within a radius of about 15$''$ is 
$M_{dyn}[M_\odot]$ $\sim$ 3.5 $\times$ 10$^{10}$.
The molecular gas contributes about 6\% to the dynamical mass in the inner 6 kpc. 
This is a typical value for a moderately gas-rich barred spiral.

\subsection{CO line ratios}

Comparison of the two CO line maps, at the same resolution and with the same 
spatial frequency sampling, can yield information about the excitation condition of the gas. 
We studied the variation in the $^{12}$CO(2-1)/$^{12}$CO(1-0) ratio in areas 
with significant emission levels in both lines, i.e. in channels with velocities 
from -180 to 220 km~s$^{-1}$. 
Within the 20.5 $''$ primary beam, the $^{12}$CO(2-1) map 
was first smoothed to the spatial resolution of the $^{12}$CO(1-0) map to assure 
that the two lines sample identical regions. The maps were then corrected for primary beam 
attenuation. 
We find a nuclear $^{12}$CO(2-1)/$^{12}$CO(1-0) line ratio close to unity and an
off-nuclear ratio around 0.2.

The values are also apparent from the maps in Fig.~3 (see caption).
The $^{12}$CO(2-1)/$^{12}$CO(1-0) 
ratio is best determined on the nucleus with a value close to unity.
The deconvolved source size in the J=2-1 line emission is 1.4$\pm$0.3~arcsec.
This results in a $^{12}$CO(2-1) line flux of 55.7 Jy~km/s~ in the 3~mm beam and a 
$^{12}$CO(2-1)/$^{12}$CO(1-0) line ratio of 0.9$\pm$0.1. 
This ratio is typical of moderately 
dense and warm (n$_{H_2}$$>$10$^4$cm$^{-3}$, T$_{kin}$$\sim$30~K),
optically thick molecular clouds if a small contribution from the cosmic microwave background 
is taken into account as well (see Eckart et al. 1990, Garc\'{\i}a-Burillo et al. 1993). 
For the weak and extended line emission along the bar, we find an 
off-nuclear  $^{12}$CO(2-1)/$^{12}$CO(1-0) 
ratio of 0.2$\pm$0.1.
This is indicative of a dominant contribution of cold (T$_{kin}$ $<$10~K) 
or sub-thermally (n$_{H_2}$ $<$10$^{-3}$cm$^{-3}$) excited molecular gas.

\subsection{Comparison with results at other wavelengths}

It is interesting to compare our results achieved through observations 
of $^{12}$CO(1-0) and $^{12}$CO(2-1) emission with observational results 
at other wavelengths. 
Laine et al. (2006) carried out 20~cm and 3.5~cm wavelength radio continuum observations with 
the VLA and compare the results with the NIR data of 
H$_{2}$ and Br$\gamma$ from Kotilainen et al. (2000).
In the radio and NIR they find 
a central point-like component and a circum-nuclear ring. 
Both source components are 
very consistent with our observations. 
The "jet", which is represented as a connection between the nucleus 
and the circumnuclear ring at 3.5 cm, 
along the major axis of the galaxy, has no counterpart in our observations.

\section{Modeling the CO distribution}

In the following section we describe the model used to analyze the data obtained with the IRAM telescopes,
then derive the rotation curve and the possible shape of the gaseous orbits.

\subsection{The model}
\label{The model}

The inner 200~pc of NGC~6574 show clear deviations from pure circular motions. 
To analyze the complex kinematics, we modeled the data with 3DRings similar 
to the approach of Krips et al.~(2005), Pott et al.~(2004), and Schinnerer et al.~(2000b; see Appendix B therein).
With 3DRings only fully symmetric structures can be modeled (i.e., no lopsidedness). The model subdivides 
the disk into many individual (circular or elliptical) orbits of molecular gas, which lie on the 
(possibly tilted) plane. With the 3DRings program we investigate the kinematical imprint of (warped) rotating gas.

\begin{figure*}
\centering
\rotatebox{0}{\resizebox{17cm}{!}{\includegraphics{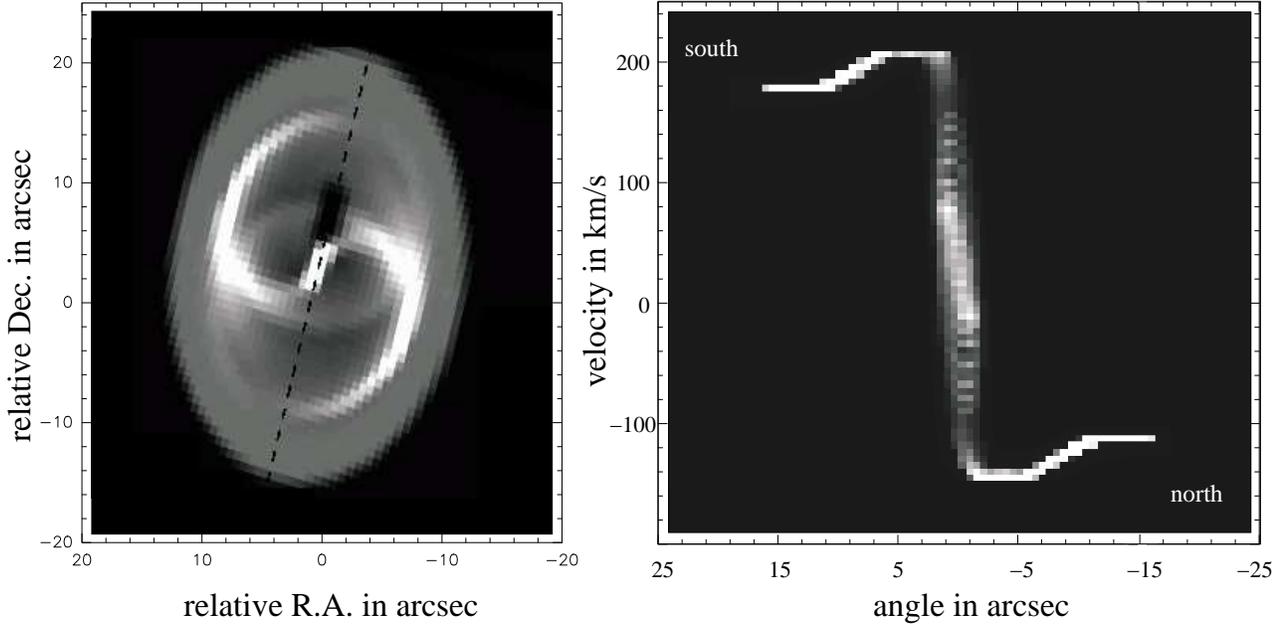}}}
\caption{Integrated intensity map and position velocity diagram along the major axis for the 
modeled  $^{12}$CO(1-0) emission.
The dashed line at PA = 165$^{\circ}$ indicates the position of the major axis of 
NGC~6574. The position-velocity
diagram was obtained as a cut made through the data cube at this position angle.
}
\label{Fig:sim}
\end{figure*}

\begin{figure}[!h]
   \centering
   \rotatebox{0}{\resizebox{7.5cm}{!}{\includegraphics{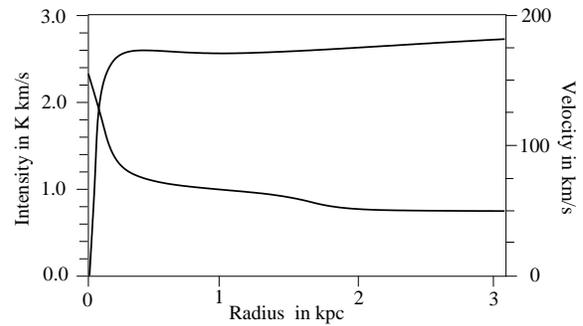}}}
   \caption{     
Intensity distribution and rotation curve as a function of radius 
as used for modeling the $^{12}$CO(1-0) line emission data.
}
   \label{Figgraph12}
\end{figure}

\begin{figure}[!h]
   \centering
   \rotatebox{0}{\resizebox{7.5cm}{!}{\includegraphics{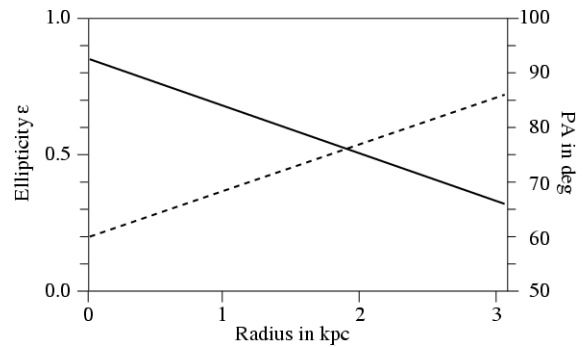}}}
   \caption{
The ellipticity $\epsilon$ (dashed line) and the position angle 
(full line) of the elliptical
bar orbits as a function of radius 
as used for modeling the $^{12}$CO(1-0) line emission data.
}
   \label{Figgraph13}
\end{figure}

We model non circular motions with 3DRings via elliptical orbits with changing position angles 
characterizing gas motions in a bar. The inclination, position angle, and shape of the rotation curve 
for the overall galaxy were held fixed. Each fitting process was started at large radii and 
successively extended toward the center. In each case we tried several start setups that all converged 
to similar (best) solutions with mean deviations from the data of less than about 10~km~s$^{-1}$ and 0.1~arcsec 
for each radius and velocity in the p-v diagrams and 10$^\circ$ in the position angle of the 
mapped structures. 
The best decompositions were achieved using a systemic velocity of 
$v_{sys}$ = 43.75 km~s$^{-1}$ in agreement with the derived position velocity diagrams 
shown in Fig.~\ref{Fig:4}~(see also spectra and channel maps in Figs.\ref{Fig:0} and \ref{Fig:1}).
The revised best-fit parameters were obtained by comparing the result with the observed 
PdBI+30m data and tuning the parameters until the match is optimal. We derived an inclination of 
45$^\circ$ $\pm$10$^\circ$ and position angle of 165$^\circ$ $\pm$10$^\circ$ for the galactic disc 
in space. In Fig.~\ref{Figgraph12} we show the intensity distribution and rotation curve 
and in Fig.~\ref{Figgraph13} the ellipticity $\epsilon$ and the position angle of the elliptical
bar orbits as a function of radius that we used to model the $^{12}$CO(1-0) line emission data.
The position angle is counted from north to east with respect to the northern part of the
inclination axis. This is represented by the dashed line in Fig.~\ref{Fig:sim}.

As a result of the modeling we find:

1) The rotation curve is basically flat until a radius of at least 1$''$ (this is the resolution 
limit of the observation and of the 3DRings model.)
This results in an enclosed mass at this radius of about 2.3$\times$10$^9$ M$_{\odot }$.

2) Starting with almost circular orbits at 20$''$ to 30$''$ radius, the ellipticity drops
to values of $\epsilon$ $\sim$0.2 ranging in position angles from
$\sim$50$^\circ$ to $\sim$90$^\circ$
approaching the bar position angle in our data (see also 
the value of  $\sim$105$^\circ$ given by Sakamoto et al. 1999ab).

3) The model requires a constant disk contribution over the entire modeled area with a 
radius of up to 30$''$. An increased central flux component starts at a radius of about 10$''$ and 
then sharply peaks at the center with a value about a factor 
of 3.0 higher than the disk component. The FWHM of the nuclear component is $\sim$1.5$''$; i.e.
it is just resolved, as shown also in the maps in Fig.~\ref{Fig:2}.

\section{Computation of the torques}
 \label{torq}

\subsection{Near infrared images}
\label{NIR}

\begin{figure}[ht]
\includegraphics[angle=-00,width=8cm]{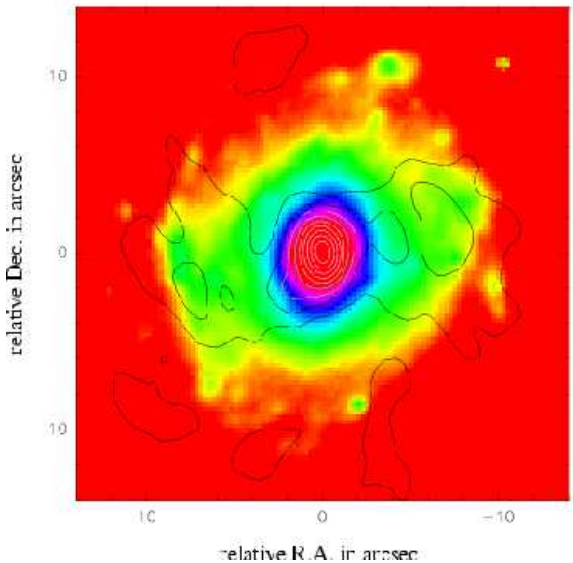}
\caption{ CO(1-0) linear contours (1.7 to 17 by 1.7 Jy/beam)
superposed on  the near-infrared H image,
from Kotilainen et al. (2000), in logarithmic levels.
The CO emission is aligned along the bar of PA = 105$^\circ$
and also possibly with the nuclear bar, inside a radius of 3$''$.}
\label{HCO}
\end{figure}

We used the near-infrared images kindly made available by J. Reunanen.
The JHK images have been obtained by Kotilainen et al. (2000)
at UKIRT under 0.6-0.7$''$ seeing and a field of view of 70$''$.
  The H band image is superposed on the CO contours
in Fig.~\ref{HCO}.  There is a very good correspondence
between the NIR image and the CO(1-0) contours, indicating that 
the molecular gas is well aligned along the bar
and nuclear bar in NGC 6574. However, a slight phase shift can be
noticed, with the gas on the leading side.
The nuclear bar is about 8$''$ (or 1.2kpc) in diameter, while
the main bar is 20$''$ (or 3.2kpc) in diameter.

\begin{figure}[ht]
\includegraphics[angle=-00,width=8cm]{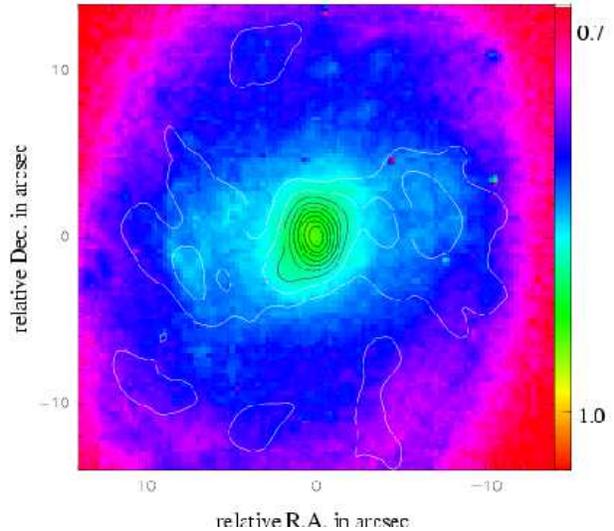}
\caption{ CO(1-0) linear contours (1.7 to 17 by 1.7 Jy/beam)
superposed on a color image in the near-infrared J-H,
from Kotilainen et al. (2000), in logarithmic levels.
The CO emission follows a spiral structure,
which winds up in a pseudo ring.}
\label{JHCO}
\end{figure}

The color J-H image is also compared with the CO emission
in Fig.~\ref{JHCO}. It reveals a spiral structure emerging 
of the main bar. 
Also, the correlation with the CO distribution is evident.
The spiral arms in the CO emission wind up in a pseudo-ring,
conspicuous in H$\alpha$ (Gonzalez-Delgado et al. 1997) and in
Br$\gamma$ (Kotilainen et al. 2000). This star-formation ring
is likely to correspond to the UHR resonance of the primary bar.
Indeed, in most studied galaxies, as shown by simulations and
confrontation with observations, the bar ends up near its UHR resonance.

\subsection{Evaluation of the gravitational potential}

As described in previous papers (e.g. Garcia-Burillo et al. 2005),
we assume that the  NIR images can give us the best approximation
for the total stellar mass distribution because it is less affected by 
dust extinction or by stellar population biases.  We here recall
the essential definitions and assumptions.

\begin{figure}[ht]
\includegraphics[angle=-90,width=8cm]{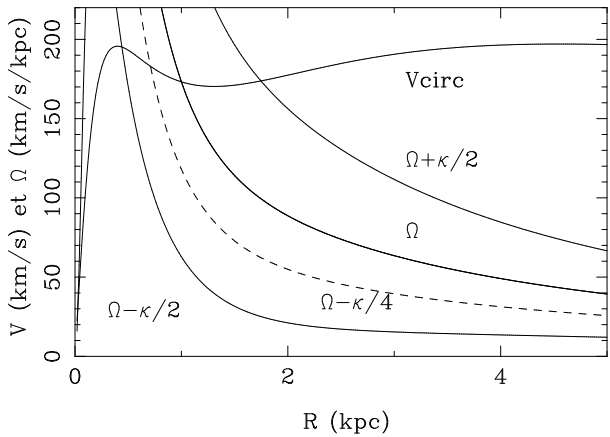}
\caption{ Rotation curve and derived frequencies $\Omega$,
$\Omega-\kappa/2$ and $\Omega+\kappa/2$, for NGC 6574. } 
\label{vcur}
\end{figure}

 The H image was first deprojected according to the
angles $PA=165^\circ$ and $i=45^\circ$.  
We did not deproject
the bulge separately, since we do not know its actual flattening, and the galaxy
- a late type (Sbc) - does not make a large contribution. 
The image is, however, completed in the vertical
dimension by assuming an isothermal plane model with a constant scale height,  
equal to $\sim$1/12th of the radial scale length of the image. The potential is
then derived by a Fourier transform method, assuming a constant mass-to-light (M/L) ratio. 
The M/L value is selected to retrieve the observed CO rotation curve. 
The axisymmetric part of the model, fitted by parametric functions, 
is then derived to find
the proper frequencies, as shown in Fig.~\ref{vcur}.
The rotation curve agrees within the uncertainties of the data
(Figs.~\ref{Fig:4})
with the rotation curve shown in Fig.~\ref{Figgraph12},
which was derived while modeling the CO data alone.

For the non-axisymmetric part, the
potential $\Phi(R,\theta)$  is then decomposed in the different m-modes:
$$
\Phi(R,\theta) = \Phi_0(R) + \sum_m \Phi_m(R) \cos (m \theta - \phi_m(R))
$$
\noindent
where $\Phi_m(R)$ and $\phi_m(R)$ represent the amplitude and phase of the m-mode.

\begin{figure}[ht]
\includegraphics[angle=-90,width=8cm]{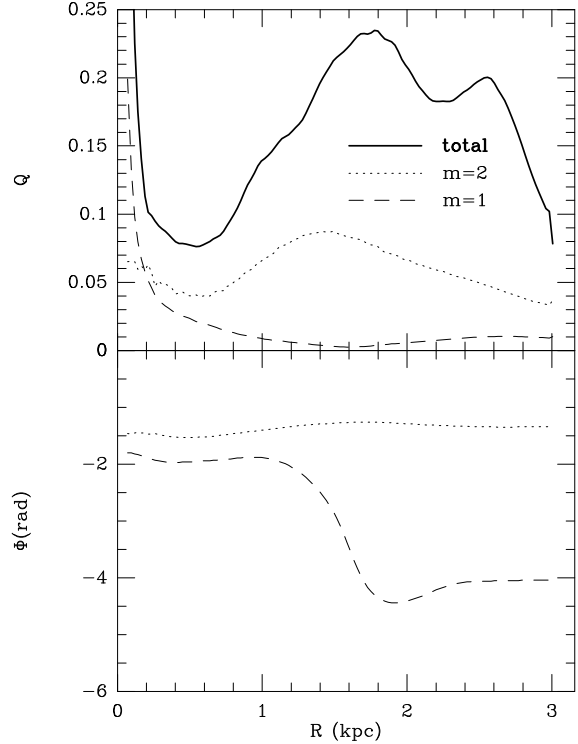}
\caption{Strengths (Q$_1$, Q$_2$, and total Q$_T$) and phases ($\phi_1$ and $\phi_2$) of the $m=1$ and $m=2$ Fourier
components of the stellar potential inside a field of 38$''$ in diameter (r $<$ 3kpc). Note the constant phase of the bar.}
\label{pot6574}
\end{figure}

The strength of the $m$-Fourier component, $Q_m(R)$, is defined as
$Q_m(R)=m \Phi_m / R | F_0(R) |$, i.e. by the ratio between tangential
and radial forces (e.g. Combes and Sanders, \cite{com81}).
The strength of the total non-axisymmetric perturbation is defined by
$$
Q_T(R) = {F_T^{max}(R) \over F_0(R)} 
$$
\noindent
where $F_T^{max}(R)$ represents the maximum amplitude of the tangential force and $F_{0}(R)$ is the mean 
axisymmetric radial force.
Fig.~\ref{pot6574}  displays these values as a function of radius for NGC 6574.
A main bar can be seen clearly, together with an asymmetry towards
the center. The lopsidedness has a correspondence in the gas
morphology described in Section 3. The bar is regular in phase, but its strength is relatively
modest.

\subsection{Evaluation of gravity  torques}

\begin{figure}[ht]
 \includegraphics[angle=-0,width=8cm]{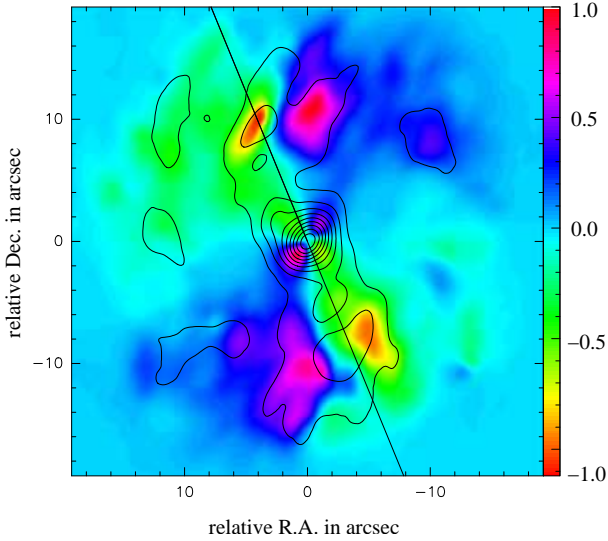}
  \caption{The CO(1--0) contours are overlaid onto the gravitational torque 
map (t(x,y)~$\times$~$\Sigma$(x,y), as defined in the text) in the center of NGC~6574. 
The deprojected torque map (grey/color scale) is 
normalized to the maximum absolute value of the torques.
The derived torques change sign as expected in a {\it butterfly} diagram, 
delineating four quadrants. The orientation of quadrants follow
the bar orientation in NGC\,6574. In this deprojected picture,
the major axis of the galaxy is oriented parallel to the abscissa. 
The inclined line reproduces the mean orientation of the bar
(PA= 105$^\circ$ on the projected image).}
  \label{torq6574}
 \end{figure}

After having calculated the forces per unit mass ($F_x$ and $F_y$) from the derivatives of $\Phi(R,\theta)$ on 
each pixel, the torques per unit mass $t(x,y)$ can be  computed by
$$
t(x,y) = x~F_y -y~F_x~~~.
$$
The torque map is oriented according to the sense of rotation in the plane of the galaxy.
The combination of the torque map and the gas density $\Sigma$(x,y) map then allows
derivation of the net effect on the gas, at each radius.
The gravitational torque map weighted by the gas surface density $t(x,y)\times
\Sigma(x,y)$, normalized to its maximum value, is shown in Fig.~\ref{torq6574}.

To estimate the radial gas flow induced by the torques, we first computed the torque 
per unit mass averaged over the azimuth, using $\Sigma(x,y)$ as the actual weighting function,i.e.
$$
t(R) = \frac{\int_\theta \Sigma(x,y)\times(x~F_y -y~F_x)}{\int_\theta \Sigma(x,y)}~~~.
$$

\begin{figure}[ht]
 \includegraphics[angle=-90,width=8cm]{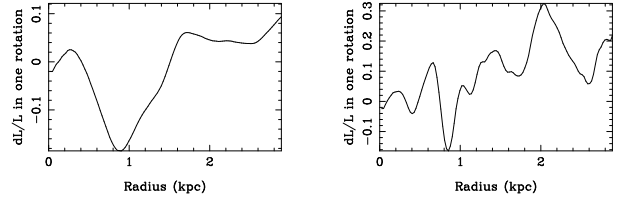}  
    \caption{The torque, or more precisely the fraction of the angular momentum transferred
from/to the gas in one rotation $dL/L$, plotted for CO(1-0) (left), and
CO(2-1) (right). The curve is noisier for the CO(2-1) line
due to a lower sensitivity at 230~GHz. Both curves, however, show negative 
torques around 1~kpc radius (more precisely from 0.4 to 1.5kpc for the CO(1-0)). }
   \label{gastor6574}
 \end{figure}

\noindent
By definition, $t(R)$ represents the time derivative of the specific angular momentum $L$ of the gas averaged 
azimuthally, i.e., $t(R)$=$dL/dt~\vert_\theta$. 
To derive non dimensional  quantities,
we normalized this variation in angular momentum per unit time
to the angular momentum at this radius and to the rotation period.
We then estimated the efficiency of the gas flow with
the average fraction of the gas specific angular momentum transferred in one rotation 
(T$_{rot}$) by the stellar potential, as a function of radius, i.e., by the function $\Delta L/L$ defined as
$$
{\Delta L\over L}=\left.{dL\over dt}~\right\vert_\theta\times \left.{1\over L}~\right\vert_\theta\times 
T_{rot}={t(R)\over L_\theta}\times T_{rot}
$$
\noindent
where $L_\theta$ is assumed to be well represented by its axisymmetric estimate, i.e., $L_\theta=R\times v_{rot}$. 
 The  $\Delta L/L$ curves for NGC6574, derived from the CO(1--0) or the CO(2--1) gas, are displayed in
Fig.~\ref{gastor6574}.

\subsection{Results and discussion}

We show in Fig.~\ref{torq6574} that the derived torques change sign following a characteristic 2D 
{\it butterfly} pattern. The CO contours clearly reveal that the majority of the gas
at r$\sim$ 0.5-1.5 kpc is offset from the bar on the leading edge where the torques
are negative. The rotation sense in the galaxy  is counterclockwise, and the spiral
structure is trailing.
Towards the center (r$<$ 400pc), however, the dominating torques are positive.
Indeed, the average value
of dL/L per rotation over 0-400pc is -0.005, and over 400-1500pc is 0.1.

The interpretation of these results can be done in the classical
scenario of angular momentum transfer in barred galaxies.
 The main bar, 1.6 kpc in radius, ends in a pseudo-ring at its 
ultra-harmonic resonance (UHR), where its pattern speed
equals $\Omega-\kappa/4$. According to Fig.~\ref{vcur},
we can then derive the probable pattern speed of $\sim$ 60 km/s/kpc.
Its co-rotation should then be around 2.8kpc. Since the rotation
curve gradient is quite strong in the center, as revealed by the CO
kinematics, there must be an ILR in the circumnuclear region,
which might correspond to the high gas concentration there,
at r$<$ 600 pc. The ILR is usually the site of high
gas concentration in a ring. 
A suggestion of such a nuclear ring
has been given in Section 3, by 
the two prominent concentrations of CO in the major axis p-v diagrams.
It is obvious that the CO emission is not peaked in the very
center, around systemic velocity, but corresponds to
an elongated nuclear ring, with highly
non-circular motions, at a radius of about 130pc.
With the pattern speed of $\sim$ 60km/s/kpc, two ILR must exist,
as suggested by the rotation curve of Fig.~11. However, this indication is
only derived from the $\Omega - \kappa/2$ curve with the epicyclic 
approximation, which is not valid for a strong bar. 
The effective curve is lower, and the ILR
might be expected between 30 and 900 pc. Moreover, if a nuclear ring exists,
it is likely to be elongated, corresponding to the range of radii
usually found in numerical simulations. 
In addition, the near infrared images give a hint of
a nuclear bar, but this is not certain. This nuclear bar is expected
from theory to decouple in such conditions, inside the nuclear ILR ring.
  
The gas flow towards the center is taken just in the act, at the present epoch.
 The flow rate is not very high, due to a bar that is relatively weak
(see Fig.~\ref{pot6574}). But the gas flow already has a visible consequence:
a high concentration of molecular gas towards the center, which must be
recent, since the center is not the location of any starburst
(no peak in Br$\gamma$ or H$\alpha$).

The maps presented here lack spatial resolution to determine 
unambiguously whether the gas is stalled at the ILR of the 
main bar, or if it could be still driven inwards due to a nuclear
bar. In either case, we conclude that we are seeing the gas inflow
fueling the center, which soon will feed the AGN nucleus.
A possibility is that the feeding is intermittent, since the gas flows
induced by gravity torques in the very center (r $<$ 400 pc) have a 
very short time scale  of  t$_{dyn} \sim$ 12~Myr.

\section{Discussion and summary}

In this paper we present the analysis of subarcsec-resolution PdBI observations 
of  molecular gas in the Seyfert~2 NGC~6574 
with a short spacing correction derived from IRAM 30m telescope observations. 
Our data complements the analysis made by Sakamoto et al. (1999ab) 
for $^{12}$CO(1-0) observations and expands the analysis with observations of 
the $^{12}$CO(2-1) line emission.
The successful descriptive modeling of the disk dynamics with elliptical bar 
orbits corroborates the hypothesis that the molecular gas is strongly 
influenced by the stellar bar-like potential
located within a more extended molecular gas disk
(Laine et al. 2006, Kotilainen et al. 2000).

Within the NUGA sample, a
whide variety of morphologies and dynamics are observed (see introduction 
and e.g. Garc\'{\i}a-Burillo et al. 2005).
With the exception of a nuclear warp,
NGC~6574 shows all of these attributes. 
But even a nuclear warp cannot be excluded as it only becomes 
apparent at the highest linear resolution (sub-100~pc), which can 
only be achieved for the closest nuclei 
(e.g. NGC~1068 and NGC~3227; e.g. Schinnerer at al. 2000a). 
In most cases, there is very little evidence
that any of the observed features can be uniquely linked to the
activity of the nucleus.  In some cases 
(NGC4826: Garc\'{\i}a-Burillo et al. 2003, NGC7217: Combes et al. 2004)
the detailed analysis of the observed source properties 
even appear to rather inhibit than to support fueling of the
nuclear region.

Contrary to the majority of NUGA galaxies studied so far,
there is  clear evidence in NGC6574 of gas inflow on
a very small scale, of a few hundred parsec, where we
find CO emission down to our spatial resolution. 
 The gas appears to accumulate at this central location,
and the inflow must be recent, since there is still no
nuclear starburst observed. 
The CO position-velocity diagrams reveal a hint 
of an elongated nuclear ring there,
which would be the ILR of the main bar. 
Our resolution does not allow us to go further to 
determine whether the gas is stalled at this ILR ring
or driven inwards to feed the AGN through
a nuclear bar.

In either case, our result is consistent with the previous finding  
that low-luminosity AGN 
(the NUGA sample includes Seyferts and LINERs) do not require 
any efficient fueling to sustain their luminosity (see e.g. Ho 2003). 
It is also unclear which processes dominate the infall
of gas into the region with radii that are beyond our current angular 
resolution i.e. linear scales of less than $\sim$100~pc.
Nuclear bars (Shlosman et al. 1989), 
lopsidedness or m = 1 instabilities 
(Garc\'{\i}a-Burillo et al. 2000), and nuclear spiral density waves 
(Englmaier \& Shlosman 2000) provide processes that 
enable the gas infall into the central 
few parsecs.
In the case of NGC~6574, both the $^{12}$CO(1-0) and the $^{12}$CO(2-1)
integrated line-intensity maps consistently show evidence of 
a weak extension toward the southeast. This may be interpreted as
indicating lopsidedness and may therefore be responsible for 
transporting molecular gas into the central 100~pc.
If, however, none of the above-mentioned mechanisms are efficient, 
then as an alternative, viscosity may support such a transport.

How viscosity may be responsible for  transporting gas toward 
the center of such disks has been shown by Duschl et al. (2000). 
The authors suggest that viscosity within these gaseous disks 
may provide an efficient AGN fueling mechanism. 
However, viscosity could also counteract low-gravity torques on the gas.
Garc\'{\i}a-Burillo et al. (2005) discuss the details of 
viscosity effects versus gravity torques to drive AGN fueling. 
The authors find that such a counteraction may occur if it 
acts on a nuclear ring of high gas-surface density contrast and a few 100~pc size. 
Whether only one of these mechanisms of different efficiency or 
a combination of them usually accounts for the gas infall is still unknown. 
Garc\'{\i}a-Burillo et al. (2005) propose an evolutionary scenario in which 
gravity torques and viscosity act in concert to produce recurrent 
episodes of activity during the typical lifetime of any galaxy. 

Alternatively, the timescales for fueling on the few 100~pc scale
and the onset of activity are so different that both cannot be 
observed simultaneously (see e.g. Combes 2004).
In that case detailed investigations of nuclear gas fueling that drives
the AGN accretion and possibly the star formation process in the 
very center of the nuclear star cluster have to resort to studies of
the molecular and atomic gas dynamics on size scales of
less than 10~pc. This could become possible with the high sensitivity and
angular resolution that will be provided by ALMA.

\begin{acknowledgements}
  Part of this work was supported by the German
 \emph{Deut\-sche For\-schungs\-ge\-mein\-schaft, DFG\/} project
 number SFB 494. We thank the IRAM stuff from the Plateau 
de Bure and from Grenoble for carrying out the 
observations and help provided during the data reduction, 
especially to J. M. Winters.
\end{acknowledgements}

\end{document}